\documentclass[conference]{IEEEtran}
\usepackage{graphicx}
\usepackage{cite}
\usepackage[bookmarks=false]{hyperref}
\usepackage{amsmath,amssymb,amsthm}
\usepackage{dsfont}
\newcommand\norm[1]{\left\lVert#1\right\rVert}
\usepackage{algorithm,algorithmic,multicol}
\usepackage{bm,upgreek}
\usepackage{color}
\usepackage{caption}
\usepackage{subcaption}
\usepackage{lettrine}
\usepackage{multirow, array, hhline}
\IEEEoverridecommandlockouts
\begin{document}
\title{Energy Efficiency Maximization in Millimeter Wave Hybrid MIMO Systems for 5G and Beyond}

\author{\IEEEauthorblockN{Aryan Kaushik, John Thompson and Evangelos Vlachos} \IEEEauthorblockA{Institute for Digital Communications, The University of Edinburgh, United Kingdom\\
Email: \{a.kaushik, j.s.thompson, e.vlachos\}@ed.ac.uk} \\
\emph{(Invited Paper)}
}

\maketitle
\begin{abstract}
At millimeter wave (mmWave) frequencies, the higher cost and power consumption of hardware components in multiple-input multiple output (MIMO) systems do not allow beamforming entirely at the baseband with a separate radio frequency (RF) chain for each antenna. In such scenarios, to enable spatial multiplexing, hybrid beamforming, which uses phase shifters to connect a fewer number of RF chains to a large number of antennas is a cost effective and energy-saving alternative. This paper describes our research on fully adaptive transceivers that adapt their behaviour on a frame-by-frame basis, so that a mmWave hybrid MIMO system always operates in the most energy efficient manner. Exhaustive search based brute force approach is computationally intensive, so we study fractional programming as a low-cost alternative to solve the problem which maximizes energy efficiency. The performance results indicate that the resulting mmWave hybrid MIMO transceiver achieves significantly improved energy efficiency results compared to the baseline cases involving analogue-only or digital-only signal processing solutions, and shows performance trade-offs with the brute force approach.
\end{abstract}

\begin{IEEEkeywords}
energy efficiency, hybrid beamforming, MIMO, millimeter wave, 5G and beyond. 
\end{IEEEkeywords}

\IEEEpeerreviewmaketitle

\section{Introduction}
Fifth generation (5G) technology is set to address the consumer demands and performance enhancements for mobile communication in 2020 and beyond \cite{networld2015}. There will be 28.5 billion networked devices and connections by 2022 \cite{cisco2016} and 8.9 billion mobile subscriptions by the end of 2024 \cite{ericsson2018}. For such large scale use of mobile devices through 5G and beyond 5G services, the communication systems would require increased capacity, high data rates, improved coverage and also reduced energy consumption. We currently use the microwave frequency spectrum for communication which is congested with a large number of consumer devices raising the demand for an unused and available spectrum. This increased demand on bandwidth and capacity can be resolved by the use of millimeter wave (mmWave) frequency spectrum which ranges from 30-300 GHz \cite{rangan2014}. This is beneficial as the larger spectral channels at mmWave would lead to higher data rates. Moreover, the large scale antenna arrays such as the multiple-input multiple-output (MIMO) systems can reduce the high path loss at mmWave frequencies \cite{heathJSTSP2016, hanCM2015}. However, it would be difficult to use one radio frequency (RF) chain per antenna leading to a least energy efficient and highly complex system. Thus, using digital beamforming which needs a dedicated RF chain per antenna is not very practical from energy efficiency (EE) and hardware complexity perspectives. To save power and reduce complexity, analogue beamforming can be used where a network of analogue phase shifters connects the antennas to a single RF chain \cite{ayachSPAWC2012}, but multi-stream and multi-user communication can not be supported.

A mmWave MIMO system with hybrid beamforming (HBF) architecture can save power and reduce hardware complexity using fewer number of RF chains than the large number of antennas, and support multi-stream communication with high spectral efficiency (SE) \cite{ayachTWC2014, aryanIET2016, payamiTVT2018, liCL2017, TsinosTCCN2019}. Such systems can also be optimized to achieve high EE gains \cite{ranziJSAC2016} but this has not been widely studied for EE maximization with low complexity. Low resolution sampling can be implemented to save power such as in \cite{aryanICC2019} we discuss EE maximization with low resolution digital-to-analogue converters (DACs) at the transmitter (TX), in \cite{aryanGCOM2019} with low resolution analogue-to-digital converters (ADCs) at the receiver (RX) and in \cite{aryanJSAC2019} with low resolution sampling at both the DACs and the ADCs. However, the existing literature mostly considers fixed number of RF chains for high SE performance \cite{ayachTWC2014, aryanIET2016, payamiTVT2018, liCL2017, TsinosTCCN2019} and RF chains consume a lot of power which increases the cost of MIMO systems \cite{bjornsonTWC2015}. Reference \cite{ranziJSAC2016} provides an exhaustive search based brute force (BF) approach where a full precoder design is evaluated for all possible combinations of RF chains, in order to select the number of RF chains that maximizes EE but this is a computationally inefficient solution. Moreover, lower complexity solutions can be implemented to design the HBF matrices than in \cite{ayachTWC2014, ranziJSAC2016}.

\begin{table}[]
    \centering
    \def\arraystretch{1.1}
    \begin{tabular}{c|c}
        \textbf{Notations} &  \textbf{Description} \\
        \hline
        $a$ & Scalar \\
        $\mathbf{a}$ & Vector \\
        $\lVert\textbf{a}\rVert_0$ & $l_0$-norm of $\textbf{a}$\\
        $\mathbf{A}$ & Matrix \\
        $|\textbf{A}|$ & Determinant of $\textbf{A}$\\
        $\mathbf{A}^{T}$ & Transpose of $\mathbf{A}$ \\ 
        $\mathbf{A}^{H}$ & Complex conjugate transpose of $\mathbf{A}$ \\ 
        $\textbf{A}^{(i)}$ & $i$-th column of $\textbf{A}$ \\
        $\Vert \mathbf{A} \Vert_{F}$ & Frobenius norm of $\mathbf{A}$\\
        $\mathcal{C}\mathcal{N} (\mathbf{a}; \mathbf{A})$ & Complex Gaussian vector; mean $\mathbf{a}$, covariance $\mathbf{A}$\\
        $\mathbb{C}^{A \times B}$ & To represent matrix of size $A \times B$ with complex entries\\
        $\mathbb{E}\{\cdot\}$ & Expectation operator \\
        $\mathbf{I}_{N}$ & Identity matrix with size $N \times N$ \\
        $\mathbb{R}^+$ & Set of positive real numbers\\
        $\mathbb{R}\{\cdot\}$ & Real part \\ 
        tr($\textbf{A}$) & Trace of $\textbf{A}$\\
        $\textbf{X} \in \mathbb{C}^{A \times B}$ & Complex-valued matrix $\mathbf{X}$ of size $A \times B$ \\
        $\textbf{X} \in \mathbb{R}^{A \times B}$ & Real-valued matrix $\mathbf{X}$ of size $A \times B$ \\
    \end{tabular}
    \caption{List of notations and their description.}
    \label{tab:notations}
\end{table}

\subsubsection*{Contribution} This paper describes different approaches to performing {\it dynamic adaptation} of a mmWave hybrid MIMO system on a frame-by-frame basis. Our idea exploits the beam training phase in the communication system to learn the propagation conditions. Based on this, we can choose to adapt the behaviour of the transceiver in order to optimize a performance metric of interest, such as EE. Maximizing EE is challenging mathematically because it is a ratio of two important parameters, namely data rate (or SE) and power. In our recent research, we use the Dinkelbach method (DM) \cite{aryanTGCN2019} to replace this ratio function by an iterative sequence of problems based on the difference of the numerator and denominator. In this work, we discuss different ways to optimize the transceivers, particularly in relation to the number of activated RF chains and the sample rate of the system. As a practical example, we present a more detailed discussion of how the Dinkelbach's approach can be used to optimize the EE and simultaneously achieve a low complexity alternative to the exhaustive search based BF approach in \cite{ranziJSAC2016}. An attractive feature of our approach is that we only need to compute the HBF matrices once, after the number of RF chains is determined by the DM based solution. 

\subsubsection*{Notations and Organization} Table I provides a list of notations used in this paper along with their description. The remainder of the paper is structured as follows: Section II describes the channel model and HBF architecture that is used in the paper. Section III describes the EE maximization problem and we describe different approaches that we have studied to address this problem. In Section IV, we discuss in more detail how the DM can be applied to select the optimal number of RF chains. Section V presents simulation results to show the performance improvements of the DM and finally Section VI presents conclusions to the paper.

\section{MmWave MIMO System with HBF}
\subsection{MmWave Channel}
We use a narrowband clustered channel model due to different channel settings at mmWave such as the number of multipaths, amplitudes, etc. \cite{heathJSTSP2016}. We consider $N_{\textrm{cl}}$ clusters with $N_{\textrm{ray}}$ paths related to each cluster and for a single user system we have $N_{\textrm{T}}$ TX antennas transmitting $N_\textrm{s}$ data streams to $N_\textrm{R}$ RX antennas. This mmWave channel can be expressed as
\begin{equation}\label{eq:channel_model}
\textbf{H} = \sum_{i=1}^{N_{\textrm{cl}}} \sum_{l=1}^{N_{\textrm{ray}}} \alpha_{il} \textbf{a}_{\textrm{R}}(\phi_{il}^{r}) \textbf{a}_{\textrm{T}}(\phi_{il}^{t})^H,
\end{equation}
where $\alpha_{il} \in \mathcal{C}\mathcal{N}(0,\sigma_{\alpha,i}^2)$ is the gain term with $\sigma_{\alpha,i}^2$ being the average power of the $i^{th}$ cluster. The vectors $\textbf{a}_{\textrm{T}}(\phi_{il}^{t})$ and $\textbf{a}_{\textrm{R}}(\phi_{il}^{r})$ denote the normalized array response vectors at the TX and the RX, respectively \cite{heathJSTSP2016}, with $\phi_{il}^{t}$ being the azimuth angles of departure and $\phi_{il}^{r}$ being the azimuth angles of arrival. We assume the transmit and receive arrays are uniform linear arrays (ULAs) of antennas, which are modelled as ideal sectored elements \cite{singh}.

\begin{figure}[t]
\centering
   	\includegraphics[width=0.515\textwidth, trim=195 75 55 125,clip]{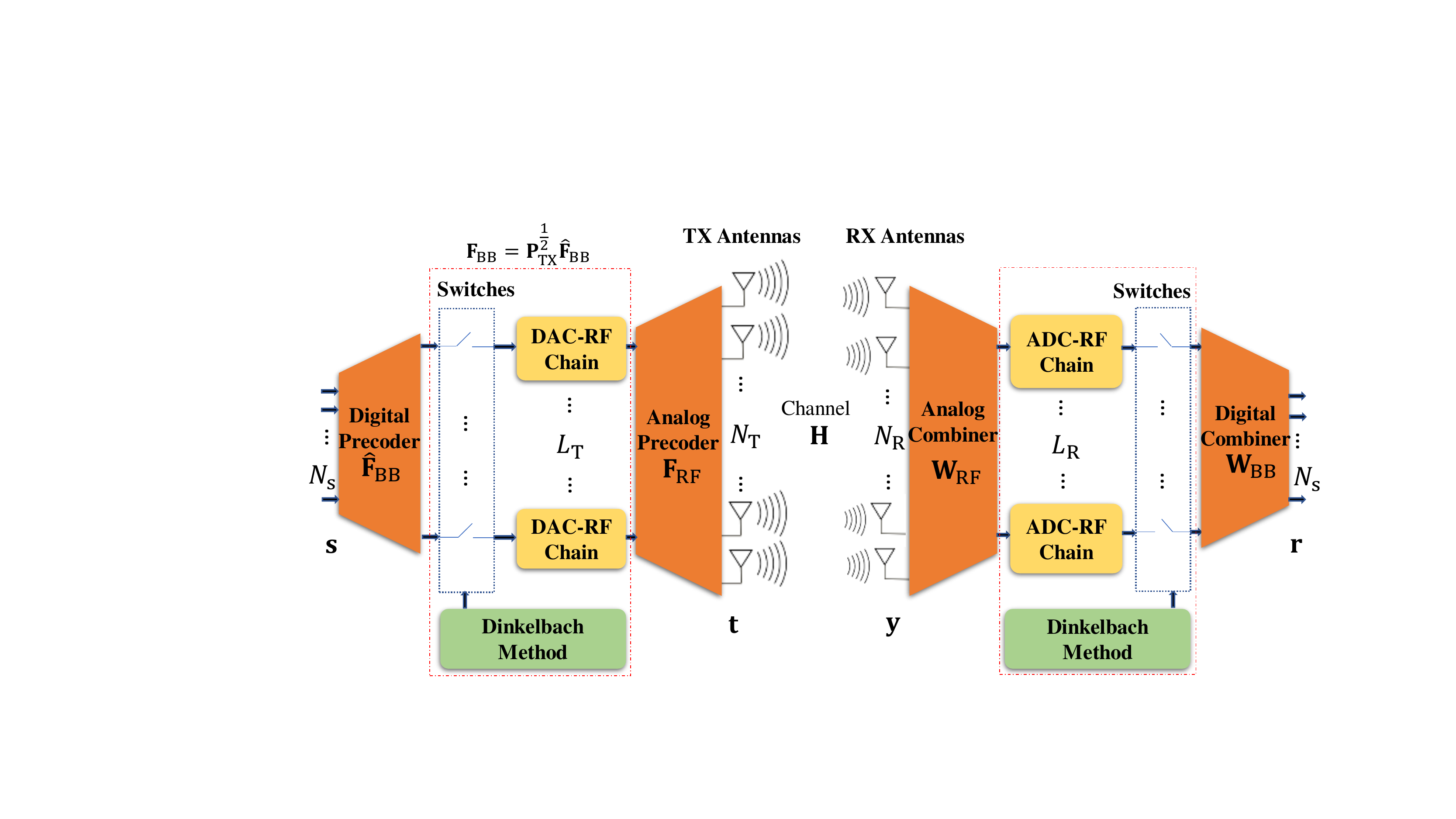}
\caption{A mmWave MIMO system with HBF architecture and the proposed DM framework.}
\end{figure}

\subsection{MIMO System with HBF Architecture}
Fig. 1 shows the system model considered in this paper where $L_\textrm{T}$ is the number of available RF chains at the TX and $L_\textrm{R}$ at the RX. Based on MIMO communication with HBF, we follow the conditions $N_\textrm{s} \leq L_\textrm{T} \leq N_\textrm{T}$ and $N_\textrm{s} \leq L_\textrm{R} \leq N_\textrm{R}$. The symbol vector $\mathbf{s} \in \mathbb{C}^{N_\textrm{s} \times 1}$ at the TX is such that $\mathbb{E}\{\mathbf{s} \mathbf{s}^H\} = \frac{1}{N_\textrm{s}}\textbf{I}_{N_\textrm{s}}$. The digital precoder matrix right before the DAC-RF chain blocks is $\mathbf{F}_\textrm{BB} \in \mathbb{C}^{L_\textrm{T} \times N_\textrm{s}} = \mathbf{P}_\textrm{TX}^{\frac{1}{2}}\hat{\mathbf{F}}_\textrm{BB}$ where $\hat{\mathbf{F}}_\textrm{BB}$ is the digital precoder matrix before the switches and $\mathbf{P}_\textrm{TX} \in \mathbb{R}^{L_\textrm{T} \times L_\textrm{T}}$ is a diagonal matrix with entries of power allocation values. We have $\textrm{tr}(\mathbf{P}_\textrm{TX}) = P_\textrm{max}$, where $P_{\textrm{max}}$ is the maximum allocated power. The entries of the analogue precoder matrix $\mathbf{F}_\textrm{RF} \in \mathbb{C}^{N_\textrm{T} \times L_\textrm{T}}$ are of constant modulus and this matrix models the phase shifting network which is only able to adjust the phase of the incoming signals, not the amplitude \cite{ayachTWC2014}. Note that the power constraint at the TX is satisfied by $\lVert\mathbf{F}_\textrm{RF}\mathbf{F}_\textrm{BB}\rVert_F^2$ = $P_{\textrm{max}}$. The matrices $\mathbf{W}_\textrm{BB} \in \mathbb{C}^{L_\textrm{R} \times N_\textrm{s}}$ and $\mathbf{W}_\textrm{RF} \in \mathbb{C}^{N_\textrm{R} \times L_\textrm{R}}$ denote the digital combiner and the analogue combiner at the RX, respectively. The analogue combiner matrix is also constant modulus.

We assume the channel state information (CSI) to be known at both the TX and the RX. Then the signal received at the RX antennas $\mathbf{y} \in \mathbb{C}^{N_\textrm{R} \times 1}$ can be written as
\begin{equation}
\textbf{y} = \mathbf{H} \mathbf{F}_\textrm{RF} \mathbf{F}_\textrm{BB} \mathbf{s} + \mathbf{n},
\end{equation}
where $\mathbf{n} \in \mathbb{C}^{N_\textrm{R} \times 1} = \mathcal{C}\mathcal{N}(0,\sigma_\textrm{n}^2)$ represents independent and identically distributed complex additive noise. After the analogue combiner and digital combiner units, the RX output signal can be expressed as
\begin{equation}
\mathbf{r} \!=\! \mathbf{W}_\textrm{BB}^H \mathbf{W}_\textrm{RF}^H \mathbf{y} \!=\! \mathbf{W}_\textrm{BB}^H \mathbf{W}_\textrm{RF}^H \mathbf{H} \mathbf{F}_\textrm{RF} \mathbf{F}_\textrm{BB} \mathbf{s} \!+\! \mathbf{W}_\textrm{BB}^H \mathbf{W}_\textrm{RF}^H \mathbf{n}.
\end{equation}
The mechanism to select only required number of RF chains $L_\textrm{T}^{opt}$ out of the available $L_\textrm{T}$ RF chains is implemented during the baseband processing. The proposed DM based solution drives this selection mechanism, which uses dynamic power allocation to decide on how many RF chains should be active during each channel realization. In the next section, we derive a fractional programming problem from the problem which maximizes EE and implement the Dinkelbach's approach to obtain the number of RF chains optimally at the TX/RX.

\section{Overview of EE Maximization}
In terms of the SE $R$ (bits/s/Hz) and the power consumption $P$ (W), the EE can be written as
\begin{equation}\label{eq:ee}
\text{EE}(\mathbf{P}_\textrm{TX}) \triangleq \frac{R(\mathbf{P}_\textrm{TX})}{P(\mathbf{P}_\textrm{TX})} \hspace{2mm}\text{(bits/Hz/J)}.
\end{equation}
In \eqref{eq:ee}, $\mathbf{P}_\textrm{TX} \in \mathcal{D}^{L_\textrm{T} \times L_\textrm{T}}$ represents a square matrix whose diagonal entries contain the transmission power of each data stream at the output of the digitally-computer precoder matrix, while all non-diagonal entries are zero. The notation $\mathcal{D}^{L_\textrm{T} \times L_\textrm{T}} \subset \mathbb{R}^{L_\textrm{T} \times L_\textrm{T}}$ represents the set of possible choices for ${L_\textrm{T} \times L_\textrm{T}}$ matrices, given the existence of a maximum transmit power constraint. 

In order to represent the selection mechanism for RF chains at the digital precoder, we consider $[\mathbf{P}_\textrm{TX}]_{kk} \in [0, P_\textrm{max}] \, \forall \, k=1,\ldots, L_\textrm{T}$. The diagonal entries of $\mathbf{P}_\textrm{TX}$ with a zero value means an open switch in the selection mechanism shown in Fig. 1. This means that the non-zero diagonal entries of the matrix $\mathbf{P}_\textrm{TX}$ determine the number of the active RF chains currently selected at the TX side, i.e., $L_\textrm{T}^{opt} = \Vert \mathbf{P}_\textrm{TX} \Vert_0$. We may achieve high SE by increasing the number of RF chains, however, it increases power consumption as well. Thus, maximizing EE in \eqref{eq:ee} given suitable constraints on the solution provides us with a practical method for selecting the TX/RX configuration with the best performance trade-off.

The optimization problem in (\ref{eq:ee}) has inspired us to study several different approaches to optimize the performance of a mmWave hybrid MIMO transceiver. As shown in Fig. 2, we deal with two phases in a single communication frame where we assume that at the start of each data frame, a beam training phase provides information to both the TX and RX about the current channel matrix $\mathbf{H}$ and there are $L_\textrm{T}$ active RF chains. Based on this knowledge it is possible to adapt the behaviour of the TX and RX before the main data communication phase, where in this paper, the DM based solution is applied to activate only required number of RF chains, i.e., $L_\textrm{T}^{opt}$, which is obtained from the solution of EE maximization problem. In the process, the HBF matrices can be designed through an Euclidean distance minimization problem \cite{ayachTWC2014} as discussed in the next section and we also propose a low complexity alternative to design the HBF matrices. Next, we discuss the approaches which we implemented to adapt the behaviour of the TX and RX in order to achieve maximum EE.   

\begin{figure}[t]
\centering
   	\includegraphics[width=0.52\textwidth, trim=200 170 150 240,clip]{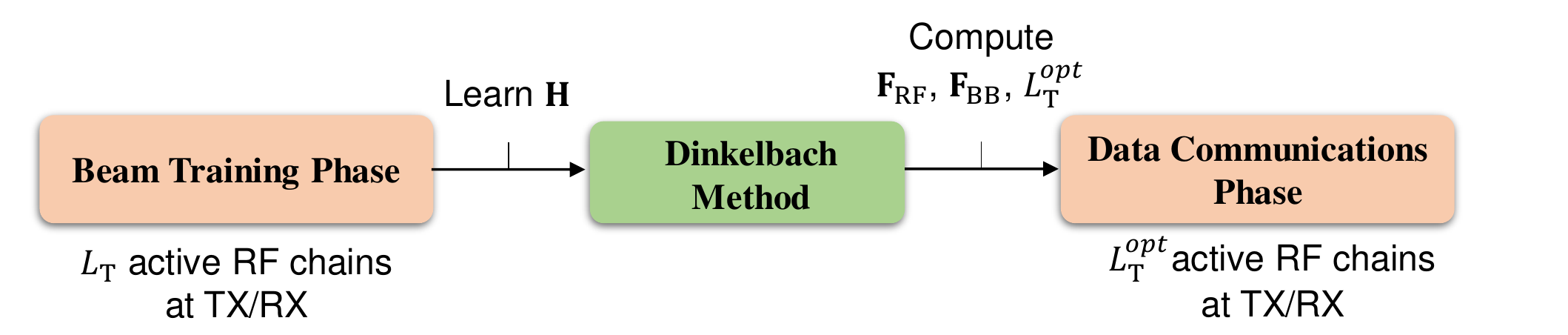}
\caption{Single communication frame with two phases process: beam training and data communications.}
\vspace{-2mm}
\end{figure}

\subsubsection{\textbf{RF Chain Selection}} In Fig. 1, the analogue precoder and the analogue combiner may connect every RF chain to every TX/RX antenna, which is termed as a {\it fully-connected} structure. Alternatively, in a structure which is termed as {\it partially-connected}, each RF chain may only be connected to a subset of all the antennas. In the latter case, we have explored an optimization technique to select the best set of RF chains for data transmission in \cite{Vlachos18}. A key feature of this approach is that we use a low signal-to-noise ratio (SNR) approximation of the data rate to simplify the optimization approach. A sparse solution for the RF chains is desired and this is obtained by minimizing the number of non-zero entries in the matrix $\mathbf{P}_\textrm{TX}$. This is achieved practically by using a technique called convex relaxation which allows the optimization to be performed efficiently. However, there is lack of research in literature dealing with the selection of RF chains. In a hardware setup, whether its fully-connected or partially-connected, when HBF is implemented on a field-programmable gate array (FPGA) chip, switching on only the needed RF chains would save a lot of power leading to an energy efficient communication system. Following that approach, in \cite{aryanTGCN2019} we consider a fully-connected structure (as shown in Fig. 1) and the Dinkelbach's approach selects only that number of RF chains which maximizes EE and the complexity is kept minimum. More details of this approach are presented in Section IV below.
\subsubsection{\textbf{Sampling Rate Selection}} A number of papers recently have shown that using limited resolution digital-to-analogue or analogue-to-digital converters in the TX or RX can improve communications efficiency \cite{Orhan15}. The reason for this is that the power consumed by a sampling device scales in an exponential manner with the number of quantization bits that are used. The limitation of using limited resolution sampling is that it can limit the overall data rate at high SNR values. However, limited resolution sampling can be particularly attractive for low or medium SNR values where the SE is lower. Reference \cite{aryanICC2019} extends the RF chain selection approach of \cite{aryanTGCN2019} to the case where the TX uses the fully-connected structure and each RF chain uses fixed resolution DACs at the TX. In that paper, a linear model is used to describe the impact of quantization, through a scaling factor and the addition of a noise term which represents the quantization noise. Similarly, the partially-connected case is with limited resolution sampling studied in \cite{Vlachos18}. We have recently extended this work to consider the joint optimization of both the HBF matrices design and the bit level resolution of each RF chain \cite{aryanGCOM2019,aryanJSAC2019}. This involves a complex model where the effect of the quantization noise on the data throughput is explicitly modelled and the bit level resolution can be adjusted to optimize the resulting EE. We introduce a novel matrix decomposition that is applied to the HBF matrices at both the TX and RX, i.e., the joint decomposition of a matrix representing analogue beamforming matrix, a second matrix modelling the impact of bit resolution on receiver noise and a third matrix that models digital baseband beamforming. Moreover, we address the joint TX-RX problem unlike in the existing literature and the optimization approach we follow requires the use of the alternating direction method of multipliers to find the best solution for both the HBF matrices and the required bit resolutions at the TX and RX in order to maximize EE. 

Next, we describe the Dinkelbach's approach for selecting the number of RF chains optimally and show how this leads to a low-cost solution to EE maximization.

\section{RF Chain Selection for Maximum EE}

\subsection{RF Chain Selection Formulation}
For MIMO with HBF and point-to-point communication, the SE $R$ given the active number of RF chains is
\begin{align}
    R(\mathbf{P}_\textrm{TX}, \mathbf{P}_\textrm{RX}) \!=\! \log \bigg\vert \mathbf{I}_{N_s} \!+\! \frac{1}{\sigma_\textrm{n}^2}\mathbf{W}_\textrm{BB}^H \mathbf{P}_\textrm{RX}^{\frac{1}{2}} \mathbf{W}_\textrm{RF}^H \mathbf{H} \mathbf{F}_\textrm{RF} \times \nonumber \\  \mathbf{P}_\textrm{TX}^{\frac{1}{2}} \hat{\mathbf{F}}_\textrm{BB} \hat{\mathbf{F}}_\textrm{BB}^H\mathbf{P}_\textrm{TX}^{\frac{1}{2}}   \mathbf{F}_\textrm{RF}^H \mathbf{H}^H \mathbf{W}_\textrm{RF} \mathbf{P}_\textrm{RX}^{\frac{1}{2}} \mathbf{W}_\textrm{BB} \bigg\vert, \label{eq:rate}
\end{align}
where the real valued $L_\textrm{T} \times L_\textrm{T}$ matrix $\mathbf{P}_\textrm{TX}$ is the diagonal matrix allocating power at the TX side. At the RX, instead we use the $L_\textrm{R} \times L_\textrm{R}$ real-valued diagonal matrix $\mathbf{P}_\textrm{RX}$ with entries from $\{0, 1\}$, since this matrix represents the activated RF chains, thus, $L_\textrm{R}^{opt} = \Vert \mathbf{P}_\textrm{RX} \Vert_0$.

Following \cite{ayachTWC2014}, we assume that $\hat{\mathbf{F}}_\textrm{BB} \hat{\mathbf{F}}_\textrm{BB}^H \approx \mathbf{I}_{L_\textrm{T}}$ and $\mathbf{W}_\textrm{BB} \mathbf{W}_\textrm{BB}^H \approx \mathbf{I}_{L_\textrm{R}}$, then the SE can be written as
\begin{align}\label{eq:updated_rate}
    R(\mathbf{P}_\textrm{TX}, \mathbf{P}_\textrm{RX}) = \log \bigg\vert \mathbf{I}_{L_\textrm{R}} + \frac{1}{\sigma_\textrm{n}^2} \mathbf{P}_\textrm{RX}^{\frac{1}{2}} \mathbf{W}_\textrm{RF}^H \mathbf{H} \mathbf{F}_\textrm{RF} \nonumber \\ \mathbf{P}_\textrm{TX}  \mathbf{F}_\textrm{RF}^H \mathbf{H}^H \mathbf{W}_\textrm{RF} \mathbf{P}_\textrm{RX}^{\frac{1}{2}} \bigg\vert.
\end{align}
The problem in \eqref{eq:updated_rate} can be simplified by considering the TX side and the RX side separately. To compute the matrix $\mathbf{P}_\textrm{TX}$ it is assumed that the RX has activated all its RF chains, so that $\mathbf{P}_\textrm{RX} = \mathbf{I}_{L_\textrm{R}}$. In that case, the SE can be expressed as
\begin{align}\label{eq:rate_tx}
    R(\mathbf{P}_\textrm{TX}) \!=\! \log \bigg\vert \mathbf{I}_{L_\textrm{R}} \!+\! \frac{1}{\sigma_\textrm{n}^2} \mathbf{W}_\textrm{RF}^H \mathbf{H} \mathbf{F}_\textrm{RF} \mathbf{P}_\textrm{TX}  \mathbf{F}_\textrm{RF}^H \mathbf{H}^H \mathbf{W}_\textrm{RF} \bigg\vert.
\end{align}

Once the matrix $\mathbf{P}_\textrm{TX}$ is obtained, the matrix $\mathbf{P}_\textrm{RX}$ can be computed via the following SE expression:
\begin{align}\label{eq:r_hn}
    R(\mathbf{P}_\textrm{RX}) \!=\! \log \bigg\vert \mathbf{I}_{L_\textrm{R}} \!+\! \frac{1}{\sigma_\textrm{n}^2} \mathbf{P}_\textrm{RX}^{\frac{1}{2}} \mathbf{W}_\textrm{RF}^H \mathbf{H} \mathbf{F}_\textrm{RF} \nonumber \\ \mathbf{P}_\textrm{TX}  \mathbf{F}_\textrm{RF}^H \mathbf{H}^H \mathbf{W}_\textrm{RF} \mathbf{P}_\textrm{RX}^{\frac{1}{2}} \bigg\vert.
\end{align}
Next, we focus on how to maximize the EE for the TX in order to select the optimal number of RF chains $L_\textrm{T}^{opt}$. The alternative of trying to solve \eqref{eq:r_hn} to maximize EE at the RX results leads to a complex integer programming optimization problem. In this paper, we will assume that the number of TX and RX spatial streams are the same, so that $L_\textrm{R}^{opt} = L_\textrm{T}^{opt}$.

Following \cite{hanCM2015}, the total consumed power $P$ for a HBF MIMO communication system can be expressed as
\begin{align}\label{eq:power}
P = \beta \textrm{tr}(\mathbf{P}_\textrm{TX}) + 2P_\textrm{CP} + N_\textrm{T} P_\textrm{T} + N_\textrm{R} P_\textrm{R} + L_\textrm{T}^{opt} \times \nonumber \\ 
(P_\textrm{RF}+N_\textrm{T} P_\textrm{PS}) + L_\textrm{R}^{opt} (P_\textrm{RF}+ N_\textrm{R} P_\textrm{PS}) \hspace{2mm} \text{(W)},
\end{align}
where the power terms $P_\textrm{CP}$, $P_\textrm{RF}$, $P_\textrm{PS}$, $P_\textrm{T}$ and $P_\textrm{R}$ represent the power required by the circuit components, the power required by each RF chain, the power required by each phase shifter, the consumed power for each antenna at the TX and that required for each RX antenna, respectively. The parameter $\beta$ is the reciprocal of amplifier efficiency.

Let us delete the subscript ``$\textrm{TX}$" from $\mathbf{P}_\textrm{TX}$ in order to write simplified expressions. Hence, the EE maximization problem in \eqref{eq:ee} can be expressed with respect to $\mathbf{P} \in \mathbb{R}^{L_{\textrm{T}} \times L_{\textrm{T}}}$ as
\begin{equation}\label{eq:ee_constraint_optimization}
\max_{\mathbf{P} \in \mathcal{D}^{L_\textrm{T} \times L_\textrm{T}}} \, \frac{R(\mathbf{P})}{P(\mathbf{P})} \, \textrm{ s. t. } \, P(\mathbf{P}) \!\leq\! P_{\textrm{max}}' \text{ \& } \, R(\mathbf{P}) \!\geq\! R_{\textrm{min}}.
\end{equation}
Note that the power constraint in \eqref{eq:ee_constraint_optimization} provides an upper limit on the power required for the HBF MIMO communication system, i.e.,
$P_\textrm{max}' = \beta P_\textrm{max} + 2P_\textrm{CP} + N_\textrm{T} P_\textrm{T} + N_\textrm{R} P_\textrm{R} + L_\textrm{T} \times (P_\textrm{RF}+N_\textrm{T} P_\textrm{PS}) + L_\textrm{R} (P_\textrm{RF}+ N_\textrm{R} P_\textrm{PS}).$ Next, we proceed with the proposed Dinkelbach's approach to obtain both the number of RF chains and the data streams optimally.

\subsection{Dinkelbach's Approach to EE Maximization}
In order to obtain a solution to \eqref{eq:ee_constraint_optimization} which is a fractional programming problem, we can implement the DM based solution. Dinkelbach's algorithm was first introduced in \cite{dinkel} and it appears to be an efficient algorithm to solve fractional problems. This is verified by the simulation results presented in Section V where we can observe that the Dinkelbach's approach achieves good performance. We can replace the EE ratio in \eqref{eq:ee_constraint_optimization} with an iterative sequence of difference-based optimizations as follows:
\begin{align}
\max_{\mathbf{P}^{(m)} \in \mathcal{D}^{L_\textrm{T} \times L_\textrm{T}}} & \left\{ R(\mathbf{P}^{(m)}) - \nu^{(m)} P(\mathbf{P}^{(m)}) \right\} \nonumber \\ &\textrm{ s. t. } \, P(\mathbf{P}) \leq P_{\textrm{max}}' \text{ and } \, R(\mathbf{P}) \geq R_{\textrm{min}}.
\end{align}
The DM involves a sequence of iterations where the constant $\nu^{(m)}$ is updated at each iteration based on the SE and power values estimated during the previous iteration which is equal to the ratio $R(\mathbf{P}^{(m-1)})/P(\mathbf{P}^{(m-1)}) \in \mathbb{R}^+$, for $m=1,2,\ldots,I_{\textrm{max}}$, where $I_{\textrm{max}}$ denotes the maximum number of iterations. In order to reduce complexity compared to the BF method, we wish to use a SE expression that does not depend explicitly on the RF and baseband processing matrices. This avoids the need to compute the HBF matrices each time the number of selected RF chains is updated. 

In order to proceed with the DM based solution, let us first update the SE and power expressions. For that, we consider channel's singular value decomposition (SVD) as $\mathbf{H} = \mathbf{U}_\textrm{H} \mathbf{\Sigma}_\textrm{H} \mathbf{V}_\textrm{H}^H$, where $\mathbf{U}_\textrm{H} \in \mathbb{C}^{N_\textrm{R} \times N_\textrm{R}}$ and $\mathbf{V}_\textrm{H} \in \mathbb{C}^{N_\textrm{T} \times N_\textrm{T}}$ are unitary matrices, and $\mathbf{\Sigma}_\textrm{H} \in \mathbb{R}^{{N_\textrm{R} \times N_\textrm{T}}}$ represents a matrix which is rectangular in nature where the diagonal entries contain the singular values of the channel matrix and all the other entries are zero. Considering the SVD of the channel, \eqref{eq:rate_tx} is written as
\begin{align}
R(\mathbf{P}) = \log \bigg\vert \mathbf{I}_{N_\textrm{R}} + \frac{1}{\sigma_\textrm{n}^2} \mathbf{W}_\textrm{RF}^H \mathbf{U}_\textrm{H} \mathbf{\Sigma}_\textrm{H} \mathbf{V}_\textrm{H}^H \mathbf{F}_\textrm{RF} \times \nonumber \\ \mathbf{P} \mathbf{F}_\textrm{RF}^H \mathbf{V}_\textrm{H} \mathbf{\Sigma}_\textrm{H}^H \mathbf{U}_\textrm{H}^H \mathbf{W}_\textrm{RF}\bigg\vert.
\end{align}
Using the approach given in \cite{ayachTWC2014}, it can be shown that $\mathbf{V}_\textrm{H}^H\mathbf{F}_\textrm{RF} \approx [ \mathbf{I}_{L_\textrm{T}} \, \mathbf{0}^T_{(N_\textrm{T}-L_\textrm{T}) \times L_\textrm{T}} ]^T  $ and $\mathbf{U}_\textrm{H}^H \mathbf{W}_\textrm{RF} \approx [ \mathbf{I}_{L_\textrm{R}} \, \mathbf{0}^T_{(N_\textrm{R}-L_\textrm{R}) \times L_\textrm{R}} ]^T$, hence,
\begin{equation}\label{eq:rate_tx_svd}
R(\mathbf{P}) = \log \bigg\vert \mathbf{I}_{N_\textrm{R}} + \frac{1}{\sigma_\textrm{n}^2} \mathbf{\bar{\Sigma}}^2 \mathbf{P} \bigg\vert,    
\end{equation}
where the $L_\textrm{R} \times L_\textrm{T}$ matrix $\mathbf{\bar{\Sigma}}$ has diagonal entries $[\mathbf{\bar{\Sigma}}]_{kk} = [\mathbf{\Sigma_\textrm{H}}]_{kk}$ for $k=1,\ldots, L_\textrm{T}$, assuming $L_\textrm{T} = L_\textrm{R}$. Again, the remaining entries of this matrix are zero. In \eqref{eq:rate_tx_svd} all of the matrices are diagonal, so it is possible to decompose the SE calculation into $L_\textrm{T}$ parallel and orthogonal channels as
\begin{align}\label{eq:rate_approximation}
    R(\mathbf{P}) \approx  \sum_{k=1}^{L_\textrm{T}} \log \left(1 +  \frac{1}{\sigma_\textrm{n}^2} [\mathbf{\bar{\Sigma}}^2]_{kk} [\mathbf{P}]_{kk}\right) \hspace{2mm} \textrm{(bits/s/Hz)}.
\end{align}
The number of available RF chains at the TX $L_\textrm{T}$ and at the RX $L_\textrm{R}$ are determined by the hardware setup of the transceiver. For the TX side, the power values in the matrix $\mathbf{P}$ can be written as
\begin{align}\label{eq:power_tx}
    P_\textrm{TX}(\mathbf{P}) &=  P_\textrm{static} + \sum_{k=1}^{L_\textrm{T}} ( \beta[\mathbf{P}]_{kk} + P_\textrm{RF} + N_\textrm{T} P_\textrm{PS}) \\
    \implies P_\textrm{TX}(\mathbf{P}) &= P_\textrm{static} + \sum_{k=1}^{L_\textrm{T}} \beta' [\mathbf{P}]_{kk}  \hspace{2mm} \textrm{(W)}, \label{eq:p_tx_over_p}
\end{align}
where the value of $P_\textrm{static} \triangleq P_\textrm{CP} + N_\textrm{T} P_\textrm{T}$ does not depend on the entries of the matrix $\mathbf{P}$ and $\beta' \triangleq \beta + \frac{ P_{\textrm{RF}} + N_{\textrm{T}} P_{\textrm{PS}}}{P_\textrm{max}}$. Simplifying  \eqref{eq:power_tx} into the form given in \eqref{eq:p_tx_over_p} is possible as $\sum_{k=1}^{L_{\textrm{T}}} [\mathbf{P}]_{kk} = \textrm{tr}(\mathbf{P}) = P_\textrm{max}$.

Following \eqref{eq:rate_approximation}-\eqref{eq:p_tx_over_p}, the $m$-th DM step can be written as
\begin{align}
& \{ \mathbf{P}^{(m)}, \nu^{(m)} \} = \textrm{arg} \max_{\mathbf{P}^{(m)} \in \mathcal{D}^{L_{\rm T} \times L_{\rm T}}} \mathcal{G}(\mathbf{P}^{(m)} \nu^{(m)}), \nonumber \\ &\textrm{ s. t. } \, P(\mathbf{P}) \leq P_{\textrm{max}}' \text{ and } \, R(\mathbf{P}) \geq R_{\textrm{min}}, \label{eq:db_step}
\end{align}
\text{where} $\mathcal{G}(\mathbf{P}^{(m)}, \nu^{(m)}) \triangleq \sum_{k=1}^{L_\textrm{T}}  \log \left(1 +  \frac{1}{\sigma_\textrm{n}^2} [\boldsymbol{\bar{\Sigma}}^2]_{kk} [\mathbf{P}^{(m)}]_{kk}\right) 
- \nu^{(m)}  \sum_{k=1}^{L_\textrm{T}} \beta' [\mathbf{P}^{(m)}]_{kk}$.
Note that \eqref{eq:db_step} is generally not convex given the constraint associated with $\mathbf{P}^{(m)}$, i.e., $\mathbf{P}^{(m)} \in \mathcal{D}^{L_\textrm{T} \times L_\textrm{T}}$. Indeed, in the case where the set $\mathcal{D}$ also contains the zero value, the problem \eqref{eq:db_step} is a mixed-integer programming one. To proceed, we alleviate this constraint on $\mathbf{P}^{(m)}$ first, so that \eqref{eq:db_step} can be solved using a standard interior-point method, e.g., using CVX \cite{cvx}. A theoretical analysis of DM convergence is presented in \cite{zappone}.

In order to explain the steps of Algorithm 1, it begins with the maximum number of RF chains $L_\textrm{T}$. Step 4 shows that we solve \eqref{eq:db_step} to update $\mathbf{P}^{(m)}$ using CVX after alleviating the constraint as mentioned above.
Then we apply the constraint again as highlighted in Step 5 of Algorithm 1. This is achieved by setting the values $\mathbf{P}^{(m)}$ to zero when they fall below the tolerance value $\epsilon_\textrm{th}$ (see Table II for $\epsilon_\textrm{th}$ value). Step 6 shows that 
counting the non-zero values of $\mathbf{P}_\textrm{th}^{(m)}$ determines the number of activated RF chains. The DM method keeps updating these values within the loop and finally computes $\lVert \mathbf{P}_\textrm{th}^{(m)}\rVert_0$ when the loop ends. Step 7 determines the SE $R(\mathbf{P}^{(m)})$ and the power $P_\textrm{TX}(\mathbf{P}^{(m)})$, and in Step 8 $\mathcal{G}(\mathbf{P}^{(m)}, \nu^{(m)})$ is computed based on 
its given expression above, where $\nu^{(m)} = R(\mathbf{P}^{(m-1)})/P(\mathbf{P}^{(m-1)}) \in \mathbb{R}^+$. Step 9 is used to update $\nu^{(m)}$ according to the current value $R(\mathbf{P}^{(m)})/P_\textrm{TX}(\mathbf{P}^{(m)})$. The loop terminates when $| \mathcal{G}(\mathbf{P}^{(m)}, \nu^{(m)})|$ is lower than the specified value $\epsilon$, which is determined empirically (see Table II for $\epsilon$ value). The number of spatial streams is then set to be equal to the optimal number of RF chains, i.e., $N_\textrm{s}$ = $L_\textrm{T}^{opt}$. 

\begin{algorithm}[t]
\begin{algorithmic}[1]
\STATE \textbf{Initialize:} $\mathbf{P}^{(0)}$, choose tolerance $\epsilon$, $L_\textrm{T}$ and set $\nu^{(0)}$ with $\mathcal{G}(\mathbf{P}^{(0)},\nu^{(0)}) \geq 0$. \\
\STATE Start Iteration Step $m = 0$.\\
\STATE \textbf{while} $| \mathcal{G}(\mathbf{P}^{(m)}, \nu^{(m)})| > \epsilon$ \textbf{do}\\
\STATE \hspace{4mm} Alleviate the constraint on $\mathbf{P}^{(m)}$ and solve \eqref{eq:db_step}. \\
\STATE \hspace{4mm} Threshold the entries of $\mathbf{P}^{(m)}$ $\rightarrow$ obtain $\mathbf{P}_\textrm{th}^{(m)}$. \\
\STATE \hspace{4mm} Count non-zero values of $\mathbf{P}_\textrm{th}^{(m)}$ $\rightarrow$ update $L_\textrm{T}^{opt}$. \\ 
\STATE \hspace{4mm} Calculate $R(\mathbf{P}^{(m)})$ and $P_\textrm{TX}(\mathbf{P}^{(m)})$ using \eqref{eq:rate_approximation}-\eqref{eq:p_tx_over_p}.\\
\STATE \hspace{4mm} Compute 
$\mathcal{G}(\mathbf{P}^{(m)}, \nu^{(m)})$.\\ 
\STATE \hspace{4mm} Update the value $\nu^{(m)}$ as $R(\mathbf{P}^{(m)})/P_\textrm{TX}(\mathbf{P}^{(m)})$.\\
\STATE \hspace{4mm} Update $m = m+1$ for next iteration.\\
\STATE \textbf{end while}\\
\STATE Compute $L_\textrm{T}^{opt}$  as the value $\lVert \mathbf{P}_\textrm{th}^{(m)}\rVert_0$.
\end{algorithmic}
\caption{Dinkelbach Method (DM)}
\end{algorithm}

Once we obtain $L_\textrm{T}^{opt}$, $L_\textrm{R}^{opt}$ ($ = L_\textrm{T}^{opt}$) and $N_\textrm{s}$, we can design the HBF matrices $\mathbf{F}_\textrm{RF}$, $\mathbf{F}_\textrm{BB}$, $\mathbf{W}_\textrm{RF}$ and $\mathbf{W}_\textrm{BB}$. We assume that as in \cite{ayachTWC2014}, the matrices  $\mathbf{F}_\textrm{RF} \mathbf{F}_\textrm{BB}$ can be designed to yield a good approximation of the fully digital precoder $\textbf{F}_\textrm{DBF}$. Note that the precoder matrix $\mathbf{F}_\textrm{DBF} = \mathbf{V}_\textrm{H1} \mathbf{P}_\textrm{TX}^{(1/2)}$ where the matrix $\mathbf{V}_\textrm{H1} \in \mathbb{C}^{N_\textrm{T}\times N_\textrm{s}}$ consists of the $N_\textrm{s}$ columns of the matrix $\mathbf{V}_\textrm{H}$ which contains the right singular eigenvectors \cite{ayachTWC2014} with $\norm{\mathbf{F}_\textrm{DBF}}_F^2 = \textrm{tr}(\mathbf{P}_\textrm{TX}) = P_\textrm{max}$. Following \cite{ayachTWC2014}, the problem to compute the hybrid precoder decomposition $\mathbf{F}_\textrm{RF} \mathbf{F}_\textrm{BB}$ through Euclidean distance minimization can be transformed to a sparse approximation problem. To solve that, we use gradient pursuit (GP) algorithm \cite{mike} which is implemented as an alternative to the most commonly used orthogonal matching pursuit (OMP) algorithm for HBF design. The GP algorithm has same performance as the OMP algorithm, but it uses only one matrix vector multiplication per iteration to avoid matrix inversion, leading to faster approximation and low complexity \cite{aryanIET2016}. At the RX, the hybrid combiner can be designed with a similar mathematical formulation as at the TX except there is no power constraint. Following the steps in \cite{ayachTWC2014}, we compute the fully digital combiner matrix $\mathbf{W}_\textrm{DBF}$ and the Euclidean distance minimization problem for the combiner design is transformed to the sparse approximation problem likewise at the TX. The sparse approximation problem at the RX can then be solved by the GP algorithm \cite{aryanIET2016} in order to obtain the hybrid combiner decomposition $\mathbf{W}_\textrm{RF}\mathbf{W}_\textrm{BB}$.

\begin{table}
    \centering
    \bgroup
\def\arraystretch{1.1}
\begin{subtable}{0.5\textwidth}
\centering
\begin{tabular}{ c|c} 
  \textbf{System Parameter} & \textbf{Value} \\
  \hline
  Number of clusters & $N_\textrm{cl} \!=\! 2$ \\
  Number of rays & $N_\textrm{ray} \!=\! 10$ \\
  Angular spread & $7.5^{\circ}$ \\
  Average power for each cluster & $\sigma_{\alpha,i} \!=\! 1$\\
  Mean angles (azimuth domain) & $60^{\circ}-120^{\circ}$ \\
  Mean angles (elevation domain) & $80^{\circ}-100^{\circ}$ \\
  Normalized system bandwidth & $1$ Hz \\
  SNR & $1/\sigma_\textrm{n}^2$ \\
  Amplifier efficiency & $1/\beta \!=\! 0.4$ \\
  Minimum desired SE in \eqref{eq:ee_constraint_optimization} &$R_{\textrm{min}} \!=\! 1$ bits/s/Hz \\
  Tolerance values & $\epsilon \!=\! {10}^{-4}$ and $\epsilon_\textrm{th} \!=\! {10}^{-6}$\\
  Number of available RF chains & $L_\textrm{T}\!=\!L_\textrm{R}\!=\!\textrm{length}\big(\textrm{eig}(\mathbf{H}\mathbf{H}^H)\big)$ \\
  Spacing between antenna elements & $d\!=\!\lambda/2$ (e.g., $\lambda\!=\!1/28$ GHz \cite{ranziJSAC2016})\\
  \end{tabular}
 \centering
 \caption{Values of the system parameters.}
\end{subtable}
\begin{subtable}{0.5\textwidth}
\centering
\begin{tabular}{ c|c}
    \textbf{Power Term} & \textbf{Value} \\
    \hline
    Power required by all circuit components & $P_\textrm{CP} \!=\! 10$ W \\
    Power required by each RF chain & $P_\textrm{RF} \!=\! 100$ mW \\ 
    Power required by each phase shifter & $P_\textrm{PS} \!=\! 10$ mW \\
    Power per TX/RX antenna element & $P_\textrm{T} \!=\!P_\textrm{R} \!=\! 100$ mW \\
    Maximum allocated power & $P_{\textrm{max}} \!=\! 1$ W \\
    \end{tabular}
    \caption{Values of the power terms in \eqref{eq:power} \cite{rappa2}.}
\end{subtable}
\egroup
\caption{Values of the system parameters and power terms used in the simulations.}
\label{tab:Table}
\end{table}

\subsubsection*{Computational Complexity} 
The computation for the DM based solution requires only $\mathcal{O}(L_\textrm{T}^{opt})$ operations per iteration. The complexity comparison with the BF approach is provided in Section V. The complexity order in computing beamforming weights for the GP algorithm is $\mathcal{O}\big((L_\textrm{T}^{opt})^3 N_\textrm{T}\big)$ and for the OMP algorithm equals $\mathcal{O}\big((L_\textrm{T}^{opt})^4 \big) + \mathcal{O}\big((L_\textrm{T}^{opt})^3$ - the GP method only makes use of  matrix multiplies at each step. This reduction in complexity comes from using a gradient computation in place of a full matrix inverse calculation. Reference \cite{aryanIET2016} provides a more detailed complexity comparison. Next, we present simulation results that verify the good performance of the proposed Dinkelbach approach.

\begin{figure}[t]
	\begin{center}
		\includegraphics[width=0.51\textwidth,trim=65 1 1 1,clip]{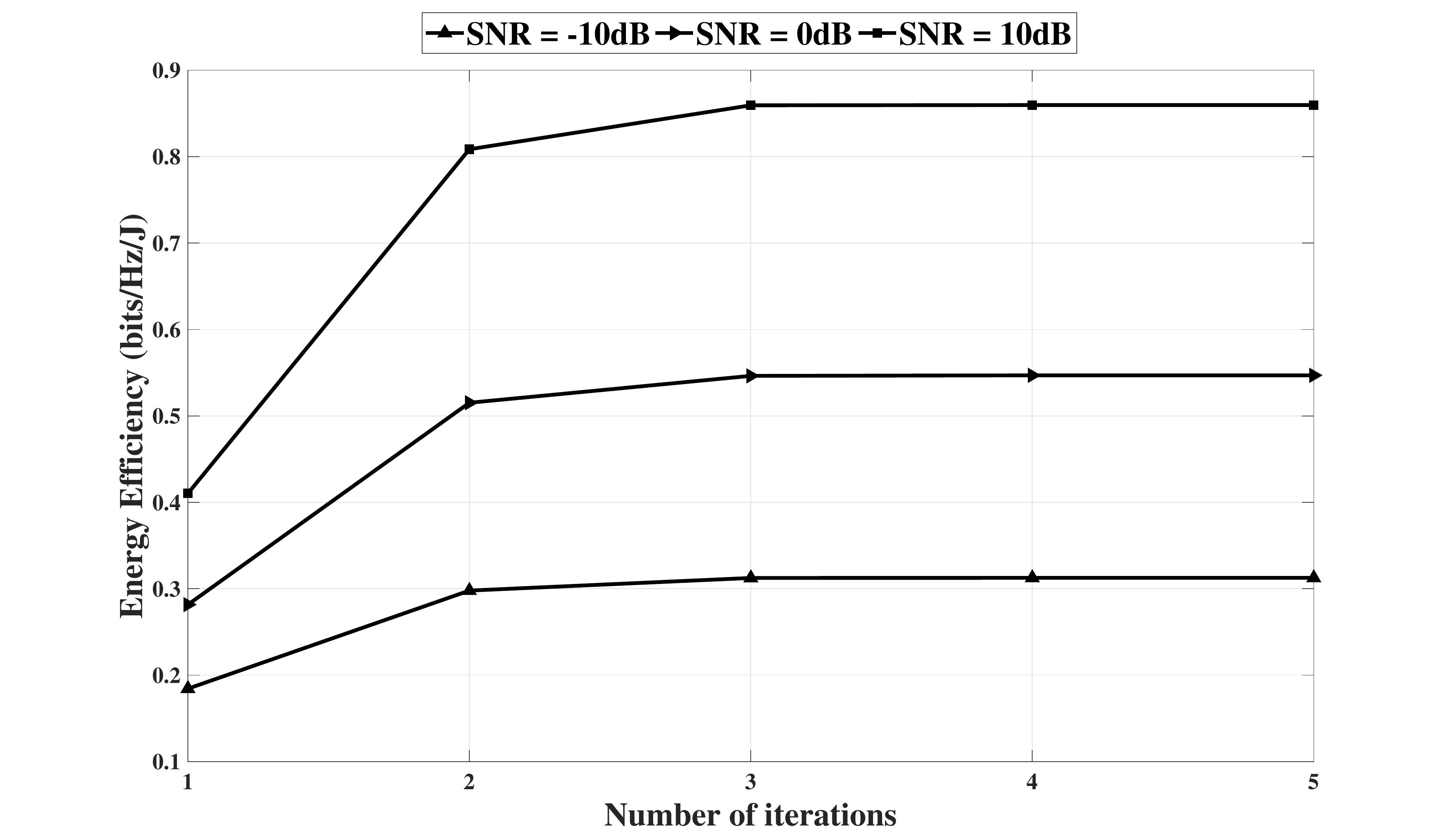}
		\caption{EE versus number of iterations at $N_\textrm{T} = 32$, $N_\textrm{R} = 8$, $N_\textrm{cl} = 2$, $N_\textrm{ray}=10$ and $P_\textrm{max} = 16$ W.}
	\end{center}
\end{figure}

\begin{figure}[t]
	\begin{center}
		\includegraphics[width=0.52\textwidth,trim=65 1 1 1,clip]{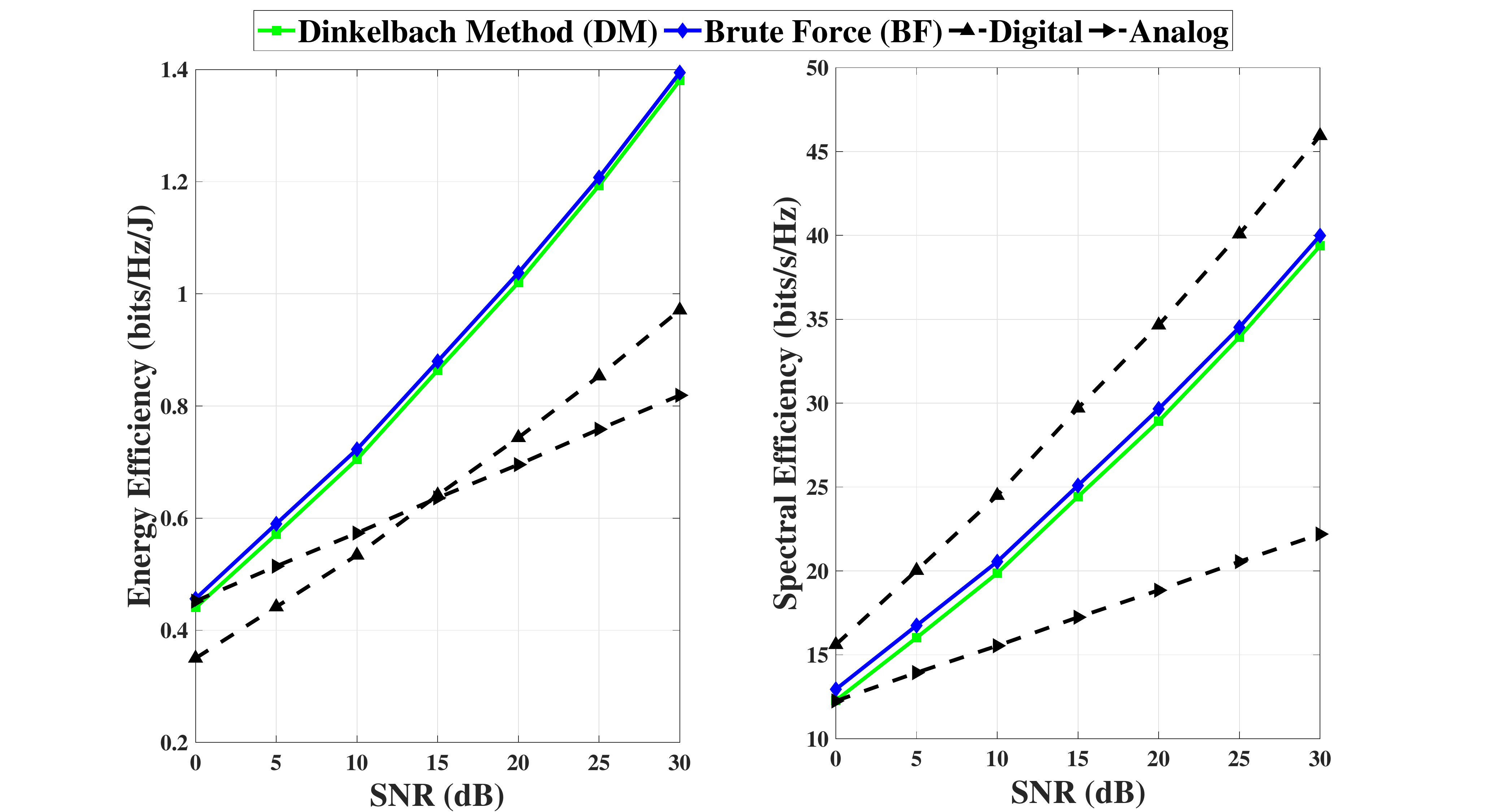}
		\caption{EE and SE versus SNR at $N_\textrm{T} = 32$, $N_\textrm{R} = 8$, $N_\textrm{cl} = 2$, $N_\textrm{ray}=10$ and $P_\textrm{max} = 1$ W.}
	\end{center}
\end{figure}

\begin{figure}[t]
	\begin{center}
		\includegraphics[width=0.52\textwidth,trim=65 1 1 1,clip]{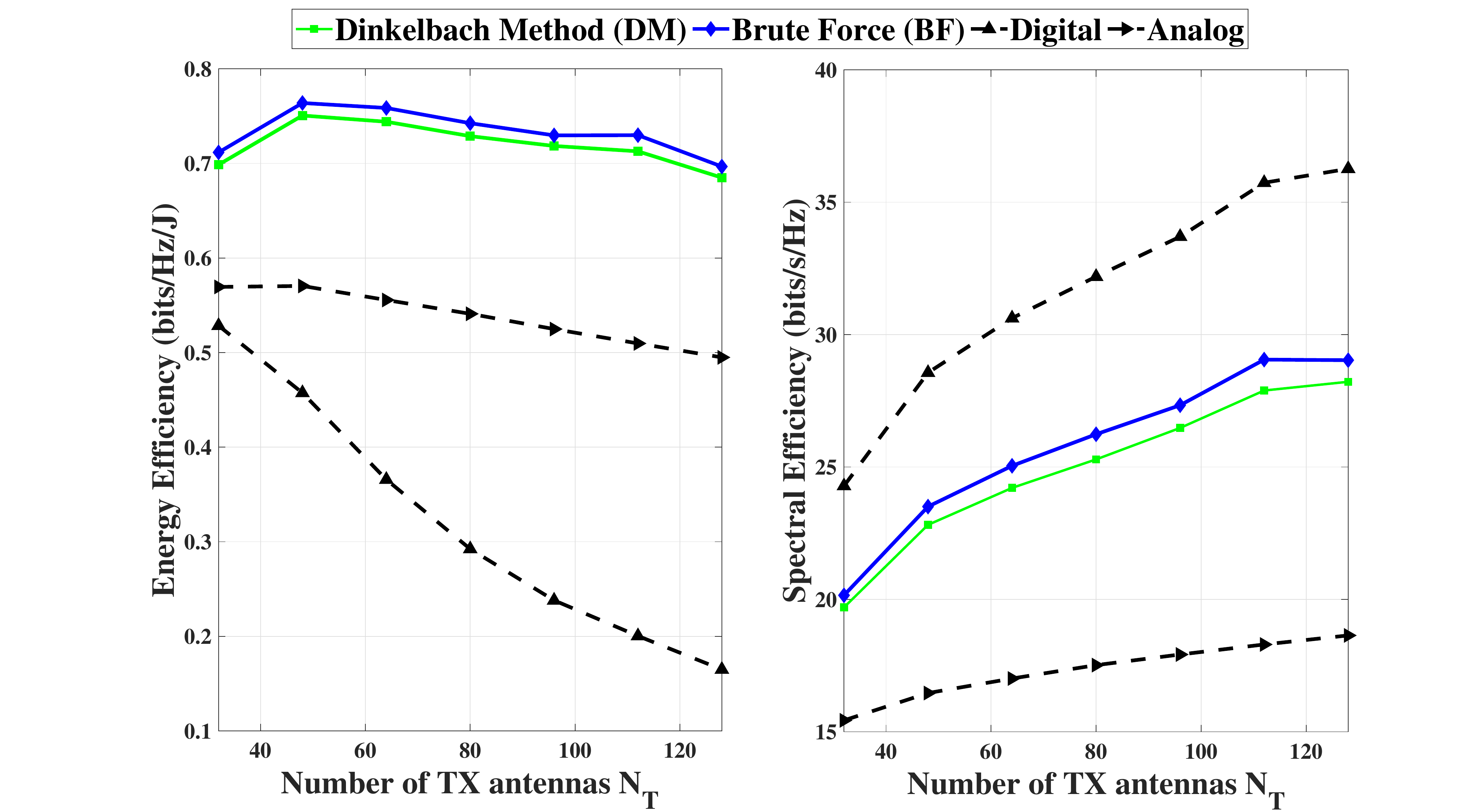}
		\caption{EE and SE versus $N_\textrm{T}$ at SNR = $10$ dB, $N_\textrm{R} = 8$, $N_\textrm{cl} = 2$, $N_\textrm{ray}=10$ and $P_\textrm{max} = 1$ W.}
	\end{center}
\end{figure}


\section{Simulation Results}
This section evaluates the performance of the proposed DM based solution and compares it with existing baseline cases. All results have been averaged over 1,000 Monte-Carlo realizations. In terms of the system setup, Table II (a) provides the values of all the system parameters and Table II (b) provides the values used in the simulations for the power terms in \eqref{eq:power}. 

For comparison with the proposed DM based solution, following baseline cases have been considered in this paper.
\subsubsection{BF Approach} The exhaustive search based approach in \cite{ranziJSAC2016}, i.e., the BF approach, at each realization (current channel realization), computes the EE performance by designing the beamforming matrices for each possible choice of activated RF chains, namely $L_\textrm{T} = \{1, 2,...,  N_\textrm{T}\}$, and then chooses the corresponding number of RF chains corresponding to the highest EE value. In contrast, the proposed DM based solution does not need to iterate for all possible number of RF chains and then find a number of RF chains which is optimal, which reduces the complexity significantly while providing high energy efficient solution. The complexity order of the BF approach is related the number of RF chains multiplied by the total number of antennas, i.e., $\mathcal{O}\big(L_\textrm{T}^{opt} N_\textrm{T} \big)$ which is larger than that of the DM based solution that only requires $\mathcal{O}(L_\textrm{T}^{opt})$ operations per iteration. In simulation, the BF and DM approaches uses the same HBF matrix computation.

\subsubsection{Digital Beamforming}
As mentioned above, the full digital beamforming baseline allocates one active RF chain for each antenna in all simulations, i.e., $L_\textrm{T} = N_\textrm{T}$ and $L_\textrm{R} = N_\textrm{R}$.  

\subsubsection{Analogue Beamforming}
In this case, analogue beamforming only implements one active RF chain , i.e., $L_\textrm{T}=L_\textrm{R} = 1$, and the HBF decomposition matrices are designed equal to phases of the first singular vectors.

Fig. 3 graphs the EE performance versus the number of iterations for SNR values of $-10$, $0$ and $10$ dB to observe convergence of the proposed DM based solution at $N_\textrm{T} = 32$, $N_\textrm{R} = 8$, $N_\textrm{cl} = 2$, $N_\textrm{ray}=10$ and $P_\textrm{max} = 16$ W. The DM based solution converges rapidly, requiring typically about two iterations to achieve an optimal solution at each channel realization. Also, the achieved EE results increase with the SNR value, for example, after $2$ iterations, the EE value at $10$ dB SNR is $\approx 0.55$ bits/Hz/J higher than that for $-10$ dB SNR and $\approx 0.3$ bits/Hz/J higher than the result for $0$ dB SNR.

Fig. 4 shows the EE and SE performance of the DM method along with the BF approach, and both the analogue and digital baseline cases versus SNR with $N_\textrm{T} = 32$, $N_\textrm{R} = 8$, $N_\textrm{cl} = 2$, $N_\textrm{ray}=10$ and $P_\textrm{max} = 1$ W. We can observe that the DM based solution has similar EE and SE performance to the BF approach, achieving a much higher EE than the digital baseline case, and higher EE and SE results compared to the analogue baseline. At an SNR value of $20$ dB, the DM based solution yield $\approx 0.2$ bits/Hz/J higher EE than the digital baseline case, and $\approx 10$ bits/s/Hz higher SE and about $0.3$ bits/Hz/J higher EE than the analogue baseline case.

Fig. 5 shows the EE and SE performance versus the number of TX antennas, $N_\textrm{T}$, plotted for an SNR of $10$ dB, $N_\textrm{R} = 8$, $N_\textrm{cl} = 2$, $N_\textrm{ray}=10$ and $P_\textrm{max} = 1$ W. It is clear that as the number of antennas increases, the EE results start to decrease for both the proposed DM based solution and the existing baseline cases. For example, at $N_\textrm{T} = 80$, the EE and SE performance of the DM based solution is similar to that of the BF method. Also, the DM based solution has $\approx 0.42$ bits/Hz/J higher EE than the digital baseline case, and $\approx 7.5$ bits/s/Hz higher SE and about $0.2$ bits/Hz/J higher EE than the analogue baseline case. 


\section{Conclusion}
This paper has discussed the concept of adaptive HBF MIMO systems that adapt their behaviour on a frame-by-frame basis to optimize EE. In particular, a DM based solution has been studied to enable fractional programming to maximize the EE of the candidate transmitter and receiver architectures in a low-cost manner. The DM method described in this paper can achieve EE and SE performance similar to the exhaustive search based BF approach, while reducing the complexity significantly. Once the number of RF chains is selected, the proposed technique needs to compute the HBF matrices only once. Further, the DM solution can also provide significantly improved EE performance when compared with the existing baseline cases, e.g., at $10$ dB SNR, it performs $\approx 20\%$ better than the digital beamforming baseline and $\approx 15\%$ better than the analogue beamforming case. Finally it is shown that the GP algorithm, which is used to compute the HBF matrices, is a faster and less complex algorithm in comparison to the state-of-the-art OMP algorithm.

\section*{Acknowledgement}
The Engineering and Physical Sciences Research Council Grant EP/P000703/1 supported this research work.
\vspace{-1.5mm}
\bibliographystyle{IEEEtran}

\end{document}